\documentclass[usenatbib]{mn2e}
\usepackage{epsf,graphics,graphicx}
\usepackage{amsmath}
\usepackage{amssymb,latexsym,mathrsfs, bm}
\usepackage{color}
\usepackage{float}
\usepackage{natbib}
\usepackage{times}
\usepackage{pdflscape}
\usepackage{longtable}
\usepackage{slashbox}
\usepackage[T1]{fontenc}
\usepackage{aecompl}

\def\eq#1{{Eq.~(\ref{#1})}}

\title[DLAs and 21 cm observations]{Modelling the cosmic neutral hydrogen from DLAs and 21 cm observations}
\author[Padmanabhan, Choudhury and Refregier]{Hamsa
Padmanabhan$^{1}$\thanks{Electronic address: hamsa@iucaa.ernet.in},
T. Roy Choudhury$^2$\thanks{Electronic address:
{tirth@ncra.tifr.res.in}},
Alexandre Refregier$^3$\thanks{Electronic address: {alexandre.refregier@phys.ethz.ch}
}\\
$^{1}$ Inter-University Centre for Astronomy and Astrophysics, Pune 411007,
India\\
$^{2}$ National Centre for Radio Astrophysics, Tata Institute of Fundamental
Research, Pune 411007, India \\
$^{3}$  Institute for Astronomy, Eidgen\"{o}ssische Technische Hochschule Zurich, Wolfgang-Pauli-Strasse 27, CH-8093 Z\"{u}rich, Switzerland}

\begin{document}
\date{ }
\maketitle

\begin{abstract}
We review the analytical prescriptions in the literature to model the 21-cm (emission line surveys/intensity mapping experiments) and Damped Lyman-Alpha (DLA) observations of neutral hydrogen (HI) in the post-reionization universe. { {While these two sets of prescriptions have typically been applied separately for the two probes, we attempt to connect these approaches to explore the consequences for the distribution and evolution of HI across redshifts. We find that a physically motivated, 21-cm based prescription, extended to account for the DLA observables provides a good fit to the majority of the available data, but cannot accommodate the recent measurement of the clustering of DLAs at $z \sim 2.3$. This highlights a tension between the DLA bias and the 21-cm measurements, unless there is a very significant change in the nature of HI-bearing systems across redshifts 0-3.}} We discuss the implications of our findings for the characteristic host halo masses of the DLAs and the power spectrum of 21-cm intensity fluctuations. 
\end{abstract}


\section{Introduction}
Understanding the distribution and evolution of neutral hydrogen in the universe provides key insights into cosmology, galaxy formation and the epoch of cosmic reionization. Two independent observational techniques are used to determine the abundance of neutral hydrogen at low and intermediate redshifts in the post-reionization universe.
At low redshifts ($z \sim 0-1$), the neutral hydrogen density distribution is studied through observations of the redshifted 21-cm emission line at radio wavelengths. 21-cm emission line surveys provide three-dimensional maps of the density and velocity fields and thus offer a tomographic probe of the universe. 
 Several galaxy surveys \citep{barnes2001, meyer2004, zwaan05, lang2003, jaffe2012, rhee13, giovanelli2005, martin10, catinella2010,lah07, lah2009} as well as intensity mapping experiments (without resolving individual clumps of galaxies, e.g. \citet{chang10, masui13, switzer13}) provide constraints on the neutral hydrogen  density ($\Omega_{\rm HI}$) and bias ($b_{\rm HI}$) parameter at low redshifts. A compilation of the currently available constraints on the neutral hydrogen density and bias parameter, and the implications for intensity mapping experiments, has been recently provided in \citet{hptrcar2015}  (hereafter Paper I). The inherent weakness of the 21-cm line transition and the limits of current radio facilities, however, hamper the detection of HI in emission at higher redshifts ($z \gtrsim 2$). Upcoming surveys, which aim to detect this weak signal at intermediate and high redshifts, include those with the the Giant Meterwave Radio Telescope \citep[][GMRT]{swarup1991}, the Meer-Karoo Array Telescope \citep[][MeerKAT]{jonas2009}, the Square Kilometre Array (SKA),\footnote{https://www.skatelescope.org} and the Canadian Hydrogen Intensity Mapping Experiment (CHIME)\footnote{http://chime.phas.ubc.ca}, among others.

At higher redshifts, our current understanding of the distribution of HI comes from the observations of Damped Lyman-Alpha systems (DLAs). DLAs are  Lyman$-\alpha$ absorption systems which have column densities $N_{\rm HI} \gtrsim 10^{20}$ cm$^{-2}$, and hence are self-shielded against the background ionizing radiation. Between the redshifts of 2 and $\sim 5$, the majority of neutral hydrogen in the universe is thought to reside in DLAs \citep{wolfe1986, lanzetta1991, storrielombardi2000, gardner1997, prochaska2005}. DLAs are also believed to be the primary reservoirs of neutral gas for the formation of stars and galaxies at lower redshifts, and hence the progenitors of today's star-forming galaxies. Studies of DLAs have primarily focussed on the Lyman-$\alpha$ absorption lines of hydrogen in the spectra of high-redshift quasars \citep{noterdaeme09, noterdaeme12, prochaska09, prochaska2005, zafar2013}.  A challenge in understanding the nature of DLAs is the identification of the host galaxies of the absorbers, the host halo masses and their properties. Imaging surveys for DLAs include those from \citet{fynbo2010, fynbo2011, fynbo2013, bouche2013, rafelski2014, fumagalli2014}, many of which point to evidence for DLAs arising in the vicinity of faint, low star-forming galaxies. Unfortunately, the presence of the bright background QSO makes the direct imaging of DLAs difficult, and thus inhibits the identification of host galaxies situated at low impact parameters. {{With large sample sizes available in current data \citep[e.g.,][]{fumagalli2014}, this caveat has lessened considerably, however, the precise understanding of the nature of DLAs still remains controversial.}}

On the theoretical front, a number of numerical simulations as well as analytical models have focussed on reproducing the observed HI content in galaxies and DLAs \citep{bagla2010, nagamine2007, pontzen2008, tescari2009, hong2010, cen2012, bird2014, dave2013, barnes2009, barnes2010, barnes2014}. Numerical methods typically involve hydrodynamical simulations with detailed modelling of star formation, feedback, self-shielding and galactic outflows. These are then compared to the available observations of $\Omega_{\rm HI}$ and $b_{\rm HI}$, and the DLA observables such as the metallicity, bias, column density distribution and velocity widths. The free parameters in the physical processes involved are thus constrained by the observational data.

Analytical approaches \citep{bagla2010, barnes2009, barnes2010, barnes2014} use prescriptions for assigning HI gas to dark matter haloes of different masses, and derive various quantities related to the 21-cm and DLA based observations.  The DLA based models typically assume a physically motivated form for the HI distribution profile \citep{barnes2010, barnes2014} and/or the cross section of DLAs \citep{barnes2009} and fix the model free parameters by fitting to the available observations. Apart from being computationally less intensive, these approaches allow direct physical interpretation of the various quantities related to the HI distribution, and their evolution across redshifts. The prescriptions thus proposed are also used together with the results of N-body \citep{guhasarkar2012} or smoothed-particle hydrodynamics (SPH) simulations \citep{navarro2014} to study the distribution of post-reionization HI. The quantities of interest in the modelling of DLAs and 21-cm observations are thus dependent upon the prescriptions for assigning HI gas to dark matter host haloes of various masses. These, in turn, are connected to the properties of the DLA hosts and the nature of the DLAs themselves. 

The 21-cm and DLA-based observations are thus independently associated with their corresponding analytical techniques in the literature. {{Here, we attempt to take into account all the available data in a common framework from both these sets of observations.}} 
 We begin by reviewing the basic features of the 21-cm and the DLA based analytical techniques in the literature. We then summarize the latest available observational data from the 21-cm and DLA-based measurements. We explore the possibility of fitting all the available data with the analytical models across redshifts, and fix the model free parameters by comparing it to the observations. Once the free parameters are fixed, the remaining observables arise as predictions of the model and can be compared directly to the data.  
 We discuss the implications of our findings for the DLA host halo masses, and the power spectrum of 21-cm intensity fluctuations to be observed in current and future experiments \citep[e.g.,][]{bull2015,santos2015}.
 
Throughout the analysis, we use the cosmological parameters $\Omega_\Lambda = 0.719, \Omega_m = 0.281, \Omega_b = 0.0462, h = 0.71, \sigma_8 = 0.8, n_s = 0.963$ which are in good agreement with most available observations, including the latest Planck results \citep{planck}. The helium fraction by mass is taken to be $Y_p = 0.24$ \citep{oliveskillman}.

\section{Theory}
In the present section, we briefly review the theoretical approaches towards modelling the observations of DLAs and neutral hydrogen intensity mapping experiments.
\subsection{21-cm based prescriptions}
The two quantities of interest associated with the neutral hydrogen intensity mapping experiments (i.e. without resolving individual galaxies) are  (a) the neutral hydrogen density parameter, $\Omega_{\rm HI}$ and (b) the bias parameter $b_{\rm HI}$. To model these properties, either analytically or through $N$-body simulations, a dark matter halo mass function $n(M, z)$ is assumed, following e.g. the Sheth-Tormen halo mass function \citep{sheth2002}. A prescription $M_{\rm HI} (M,z)$ is used to assign HI gas (of mass $M_{\rm HI}$) to the halo (of mass $M$). 
The neutral hydrogen density parameter $\Omega_{\rm HI}$ is then  defined from the prescription as:
\begin{equation}
 \Omega_{\rm HI} (z) = \frac{1}{\rho_{c,0}} \int_0^{\infty} n(M, z) M_{\rm HI} (M,z) dM
 \label{omegaHI}
\end{equation} 
where $n(M,z)$ denotes the distribution of the dark matter halos and $\rho_{c,0}$ is the critical density of the universe at redshift 0.
Given the prescription for $M_{\rm HI} (M,z)$, the HI bias may be calculated as:
\begin{equation}
b_{\rm HI} (z) = \frac{\int_{0}^{\infty} dM n(M,z) b (M,z) M_{\rm HI} (M,z)}{\int_{0}^{\infty} dM n(M,z) M_{\rm HI} (M,z)}
\label{biasHI}
\end{equation}
where the dark matter halo bias $b(M,z)$ is given, for example, following \citet{scoccimarro2001}.

In surveys of 21-cm emission from galaxies, the lower limits in the integrals in \eq{biasHI} are fixed to the halo mass $M_{\rm min}$ corresponding to the minimum HI mass observable by the survey. This leads to the expression for the HI bias measured from galaxy surveys \footnote{ \eq{biasHIgal} makes the  approximation that the selection function of the HI galaxies can be modelled by a combination of HI mass weighting and a low-mass cutoff.}:
\begin{equation}
b_{\rm HI,gal} (z) = \frac{\int_{M_{\rm min}}^{\infty} dM n(M,z) b (M,z) M_{\rm HI} (M,z)}{\int_{M_{\rm min}}^{\infty} dM n(M,z) M_{\rm HI} (M,z)}
\label{biasHIgal}
\end{equation}

\subsection{DLA based prescriptions}
 In the case of Damped Lyman Alpha system (DLAs) observations, the primary observable is the distribution function of neutral hydrogen in DLAs, i.e. $f_{\rm HI} (N_{\rm HI})$ where $N_{\rm HI}$ is the column density of the DLAs. From the observations of $f_{\rm HI} (N_{\rm HI})$, the density parameter of neutral hydrogen in DLAs, $\Omega_{\rm DLA}$ and $dN/dX$, the incidence rate of the DLAs per unit comoving absorption path length, may be inferred. Recently, the bias parameter of DLAs, $b_{\rm DLA}$ has been estimated at redshift $z \sim 2.3$ by cross-correlation studies with the Lyman-$\alpha$ forest \citep{fontribera2012}. To explain the measured values of these observables, the DLA may be modelled as an absorbing cloud in a host dark matter halo of mass $M$, with a neutral hydrogen density profile $\rho_{\rm HI}(r)$ as a function of $r$.

The column density of DLAs is then calculated using the relation:
\begin{equation}
 N_{\rm HI}(s) = \frac{2}{m_H} \int_0^{\sqrt{R_v(M)^2 - s^2}} \rho_{\rm HI} (r = \sqrt{s^2 + l^2}) \ dl 
 \label{coldenss}
\end{equation} 
where $m_H$ is the mass of the hydrogen atom, $R_v(M)$ is the virial radius associated with a halo of mass $M$ and $s$ is the impact parameter of a line-of-sight through the DLA. 
The cross-section $\sigma_{\rm DLA}$ is defined as $\sigma_{\rm DLA} = \pi s_*^2$ where $s_*$ is the root of the equation $N_{\rm HI}(s_*) = 10^{20.3}$ cm$^{-2}$. 
The DLA bias $b_{\rm DLA}$ is defined by:
\begin{equation}
 b_{\rm DLA} (z) =  \frac{\int_{0}^{\infty} dM n (M,z) b(M,z) \sigma_{\rm DLA} (M,z)}{\int_{0}^{\infty} dM n (M,z) \sigma_{\rm DLA} (M,z)}.
\end{equation} 
The incidence $dN/dX$ is calculated as:
\begin{equation}
 \frac{dN}{dX} = \frac{c}{H_0} \int_0^{\infty} n(M,z) \sigma_{\rm DLA}(M,z) \ dM
 \label{dndxdef}
\end{equation} 
The column density distribution $f_{\rm HI}(N_{\rm HI}, z)$ is given by:
\begin{equation}
 f(N_{\rm HI}, z) = \frac{c}{H_0} \int_0^{\infty} n(M,z) \left|\frac{d \sigma}{d N_{\rm HI}} (M,z) \right| \ dM 
 \label{coldensdef}
\end{equation} 
where the $d \sigma/d N_{\rm HI} =  2 \pi \ s \ ds/d N_{\rm HI}$, with $N_{\rm HI} (s)$ defined as in \eq{coldenss}.

Finally, the density parameter for DLAs, $\Omega_{\rm DLA}$ is calculated as:
\begin{equation}
 \Omega_{\rm DLA}(N_{\rm HI}, z)  = \frac{m_H H_0}{c \rho_{c,0}} \int_{10^{20.3}}^{\infty} f_{\rm HI}(N_{\rm HI}, z) N_{\rm HI} d N_{\rm HI}
\end{equation} 

Alternatively, the cross section $\sigma_{\rm DLA} (M)$ itself is modelled using a functional form, and the DLA quantities may be directly calculated from $\sigma_{\rm DLA}$.

\section{Data}

In the present section, we compile the currently available constraints from the 21-cm and DLA observations (a detailed summary is available in Paper I):

(a) Constraints on $\Omega_{\rm HI}$ from 21-cm galaxy surveys include those from   \citet[][HIPASS and the Parkes observations of the SGP field at $z \sim 0.03$ and 0.1]{delhaize13}, \citet[][WSRT 21-cm emission at $z = 0.1$ and 0.2]{rhee13}, \citet[][ALFALFA survey observations]{martin10}, \citet[][AUDS survey at $z \sim 0.125$]{freudling11}, and
\citet[][co-added observations from the GMRT at $z = 0.24$]{lah07}. 

(b) Joint constraints on the product $\Omega_{\rm HI} b_{\rm HI}$ from 21-cm intensity mapping experiments at $z \sim 0.8$ come from  \citet{chang10}, \citet{masui13} and \citet{switzer13}. 

(c) Constraints on the neutral mass fraction in DLAs, $\Omega_{\rm DLA}$ at $z \sim 0$ are available from \citet[][observations of HI distribution in M31, M33 and the Large Magellanic Cloud (LMC)]{braun2012},\citet[][HIPASS catalogue, $z \sim 0$]{zwaan05}, and at $0.5 < z < 5$ from the observations of \citet{rao06}, \citet{prochaska09}, \citet{noterdaeme09}, \citet{noterdaeme12} and \citet{zafar2013}.

(d) Constraints on the bias parameter $b_{\rm HI}$ at $z \sim 0$ are available from the ALFALFA survey \citep{martin12}.

(e) Constraints on the DLA incidence $dN/dX$ at $z \sim 0$ are available from \citet{braun2012}, \citet{zwaan2005a}, and at $z \sim 1$ from \citet{rao06}. \citet{zafar2013} compiles the currently available constraints on the DLA incidence $dN/dX$ over $1.5 < z < 4$.

(f) The HI column density distribution $f_{\rm HI}$ at redshift $z \sim 2.5$ comes from the observations of \citet{noterdaeme12}. The column density distribution at redshift $z \sim 0$ is available from the observations of \citet{zwaan2005a} and at $z \sim 1$ from \citet{rao06}.
 
(g)  \citet{fontribera2012} provide a measurement of the bias parameter $b_{\rm DLA}$ of DLAs at redshift $z \sim 2.3$.

\section{Models}
We now attempt to model the neutral hydrogen distribution by considering both DLAs and 21-cm intensity mapping measurements. To begin with, we describe some of the models available in the literature to assign HI to dark matter haloes and compute the relevant quantities in the 21-cm and DLA based prescriptions.

\begin{enumerate} 
 \item 21-cm based: \citet{bagla2010} explore several prescriptions for assigning HI gas to dark matter haloes to be used with the results of N-body simulations. In their simplest model,  $M_{\rm HI} (M)$ is modelled as $M_{\rm HI} = fM$ between limits $M_{\rm min}$ and $M_{\rm max}$ that correspond to virial velocities of 30 km/s and 200 km/s respectively.  The value of $f$ is chosen to match the observations of $\Omega_{\rm HI}$. 
 
\item DLA based: 
   \citet{barnes2009, barnes2010} develop analytical models to explain various observed properties of the DLAs at $z \sim 3$, such as their column density distribution and velocity width distribution. \citet{barnes2014} jointly model the column density distribution, the velocity width distribution of associated low-ionization metal absorption and the observed bias parameter of DLAs at $z \sim 2.3$.
   
   In \citet{barnes2009}, the cross-section $\sigma_{\rm DLA} (M)$ is modelled using the functional form:
   \begin{equation}
\sigma_{\rm DLA}(M) = \pi r_0^2 \left(\frac{v_c}{200 {\rm km/s}} \right)^{\beta} \exp \left[-\left(\frac{v_{c,0}}{v_c}\right)^{a}\right]                                                                                                                                                                                                                              \end{equation} 
with $\beta = 2.5$, $v_c(M)$ is the virial velocity for a halo of mass $M$,  and the free parameters $a, r_0$ and $v_{c,0}$ are fixed by comparing the model predictions to the available observations.

   The above model is mentioned only for completeness  and we do not consider it further for fitting the data. The other  two DLA-based  models  \citep{barnes2010, barnes2014} assign HI to dark matter matter haloes according to mass prescriptions, which enables comparison to the 21-cm based model. We hence consider these two DLA-based models  for the remainder of the text. Both the models use the prescription for $M_{\rm HI} (M)$ to be given by:
  \begin{equation}
    M_{\rm HI} (M) = \alpha M f_{H,c} \exp\left[-\left(\frac{v_{c,0}}{v_c(M)}\right)^3\right]
   \end{equation} 
   where $\alpha$ is the neutral fraction of HI in the halo (relative to cosmic),  $f_{H,c} = (1 - Y_p) \Omega_b/\Omega_m$ is the cosmic hydrogen fraction and $Y_p$ is the cosmological helium fraction by mass, and the parameter $v_{c,0}$ is a free parameter fixed by matching to the observations. The best-fit values are $v_{c,0} = 50$ km/s in \citet{barnes2010} and $v_{c,0} = 90$ km/s in \citet{barnes2014}. In both models, the gas radial distribution profile is modelled as an altered NFW profile:
   \begin{equation}
    \rho_{\rm HI}(r) = \frac{\rho_0 r_s^3}{(r + 0.75 r_s) (r+r_s)^2}
    \label{rhodef}
   \end{equation} 
   where $r_s$ is the scale radius, defined as $r_s = R_v(M)/c(M,z)$ with $R_v (M)$ being the virial radius:
   \begin{equation}
 R_v (M) = 46.1 \ {\rm{kpc}} \  \left(\frac{\Delta_v \Omega_m h^2}{24.4} \right)^{-1/3} \left(\frac{1+z}{3.3} \right)^{-1} \left(\frac{M}{10^{11} M_{\odot}} \right)^{1/3}
\end{equation} 
with $\Delta_v = 18 \pi^2 + 82 d - 39 d^2$ and $d = \Omega_m(z) - 1 =    \Omega_m (1+z)^3/(\Omega_m (1+z)^3 + \Omega_{\Lambda}) - 1$ \citep{bryan1998}.
The parameter $c$ is the halo concentration approximated by:
\begin{equation}
 c(M,z) = c_{\rm HI} \left(\frac{M}{10^{11} M_{\odot}} \right)^{-0.109} \left(\frac{4}{1+z} \right)
\end{equation} 
where $c_{\rm HI}$ is the concentration parameter for the HI, which is analogous to the dark matter halo concentration $c_0 = 3.4$ in \citet{maccio2007}. Typically, one finds that $c_{\rm HI}$ is larger than $c_0$.
The normalization $\rho_0$ in \eq{rhodef} is determined by the condition that:
\begin{equation}
 \int_0^{R_v(M)} 4 \pi r^2 \rho_{\rm HI}(r) dr = M_{\rm HI} (M)
\end{equation} 
   The best-fit value of the parameter $c_{\rm HI}$ is found to be $c_{\rm HI} = 25$ in both DLA based models. 
   
   \end{enumerate}
   
\begin{figure}
   \begin{center}
\includegraphics[scale = 0.45]{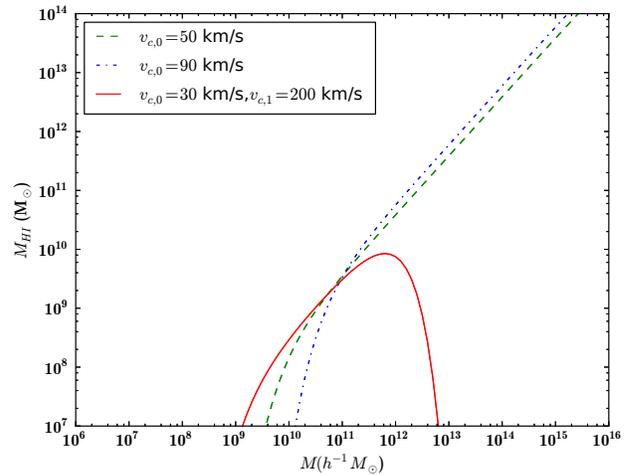} 
   \end{center}
\caption{Prescriptions for the 21-cm based and the two DLA based models (plotted here at $z = 2.3$). The difference between the prescriptions arises from the range of mass values probed.}
   \label{fig:prescriptions}
  \end{figure}

   We now attempt to combine the above three prescriptions into similar functional forms. To do this, we introduce exponential lower and upper cutoffs in the 21-cm based mass prescription, of 30 and 200 km/s respectively:
     \begin{equation}
    M_{\rm HI} (M) = \alpha f_{\rm {H,c}} M \exp\left[-\left(\frac{v_{c,0}}{v_c(M)}\right)^3\right] \exp\left[-\left(\frac{v_c(M)}{v_{c,1}}\right)^3\right]
    \label{prescrip}
   \end{equation}
   where $v_{c,0} = 30$ km/s,  $v_{c,1} = 200$ km/s, and $\alpha$ is an overall normalization parameter which denotes the fraction of HI in the halo relative to cosmic. The functional form adopted in \eq{prescrip} also applies to the two DLA based prescriptions, with $v_{c,1} = \infty$\footnote{In practice, we may adopt a very high ($\sim 10000$ km/s) upper cutoff for these models. The decline of the mass function at the high mass end provides a natural high mass cutoff to these prescriptions.} and $v_{c,0} = 50$ km/s and 90 km/s, respectively. Hence, we may parametrize the three prescriptions by their cutoff velocities. These are plotted in Fig. \ref{fig:prescriptions}.
   
   The quantities $\alpha, c_{\rm HI}, v_{c,0}$ and $v_{c,1}$ have direct physical implications in the calculations of the relevant observables involved in estimating the 21-cm HI power spectrum. $\alpha$ provides the overall normalization (the HI fraction in the halo relative to cosmic) and $c_{\rm HI}$ describes the concentration of HI relative to the underlying dark matter distribution. 
  These two parameters may be fixed by comparing to the observations of $dN/dX$ and $f_{\rm HI}$.  The two cutoffs, $v_{c,0}$ and $v_{c,1}$ select the range in the mass function that contributes significantly to the bias $b_{\rm HI}$ (at low redshifts) or $b_{\rm DLA}$ (at higher redshifts). Note that in the absence of the cutoffs (i.e. if $M_{\rm HI} \propto M$ for all $M$), we would have:
 \begin{equation}
  b_{\rm HI} (z) = \frac{\int_{0}^{\infty} dM n(M,z) b (M,z) M}{\int_{0}^{\infty} dM n(M,z) M} = 1
 \end{equation} 
 which follows from the results in \citet{seljak2000,refregier2002}. Hence, the lower and upper cutoffs in the HI mass prescription are directly related to the calculated bias.

   It can be seen from Fig. \ref{fig:prescriptions} that the 21-cm and DLA based models are very different with respect to the choice of host halo masses for assignment of neutral hydrogen. The results of simulations disfavour the assignment of HI gas to  halos with virial velocities smaller than $30$ km/s (or halo masses below $10^{9} M_{\odot}$ at $z \sim 2.3$). Also, haloes of masses corresponding to virial velocities greater than 200 km/s ($M \sim  10^{12} M_{\odot}$ at $z \sim 2.3$) are not expected to host HI \citep{pontzen2008}. We, therefore, attempt here to extend the 21-cm based prescription to also account for the available DLA observables.

   \begin{figure}
   \begin{center}
    \includegraphics[scale = 0.3, angle = 90]{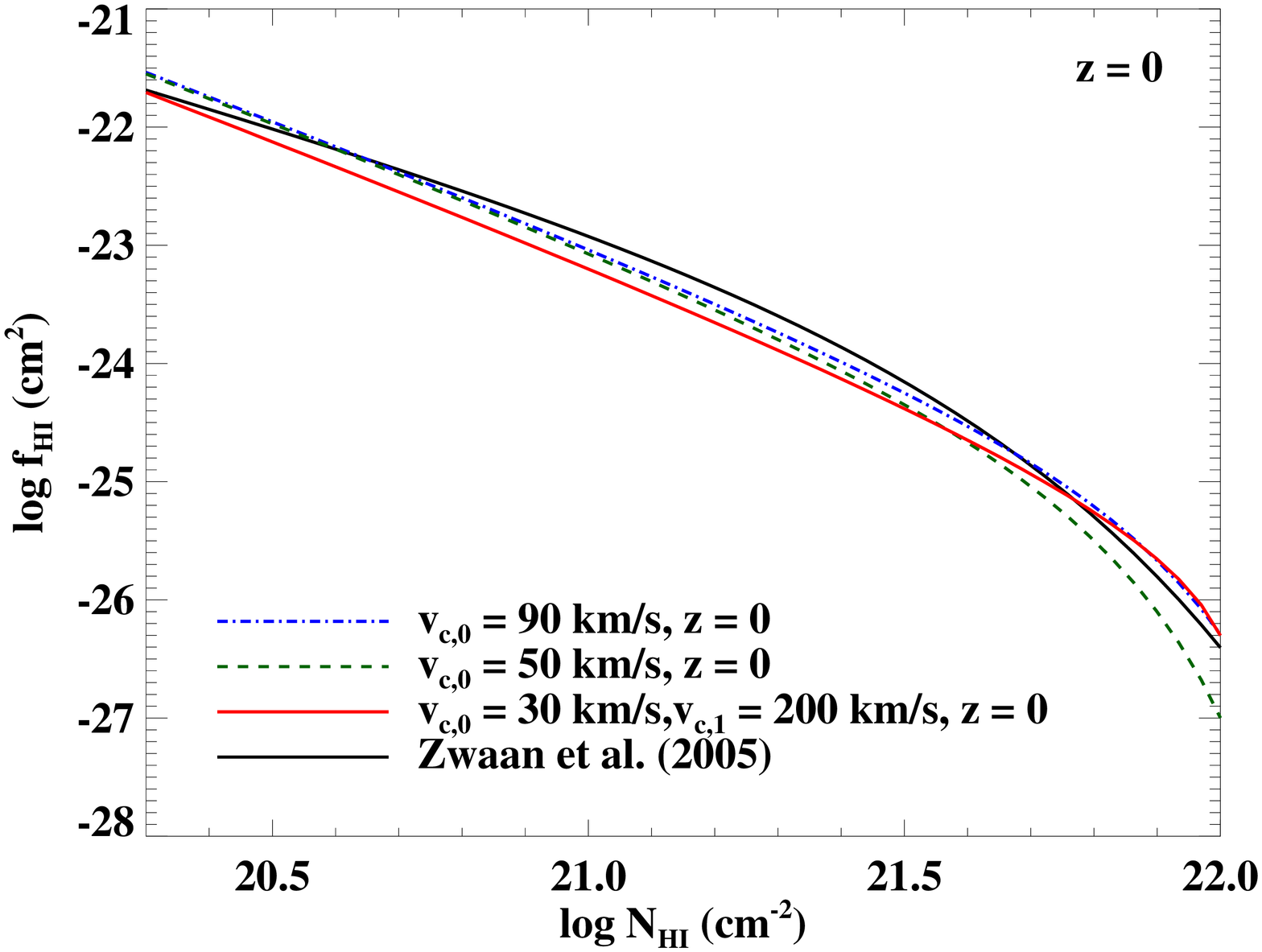} \includegraphics[scale = 0.3, angle = 90]{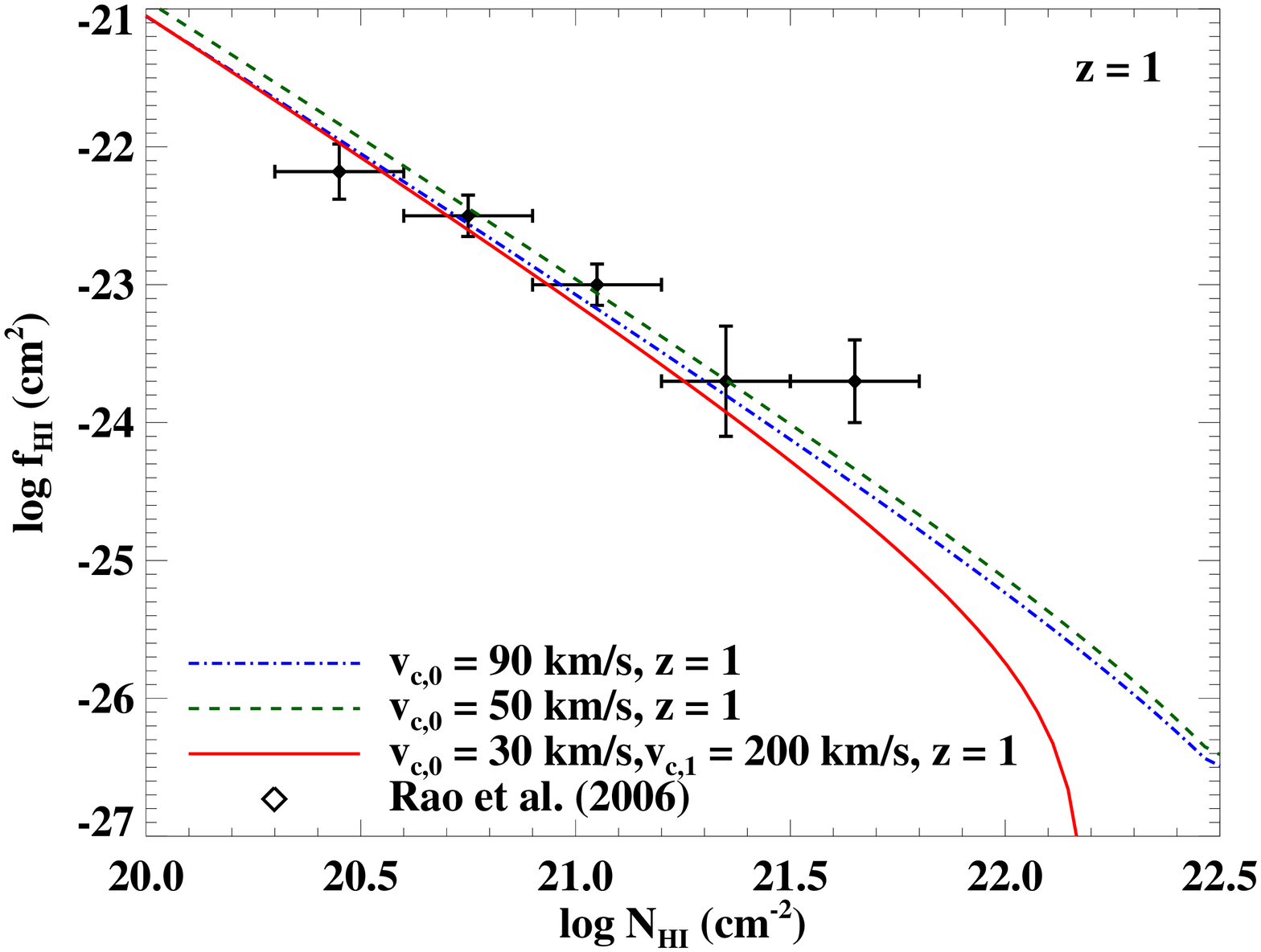} \includegraphics[scale = 0.3, angle = 90]{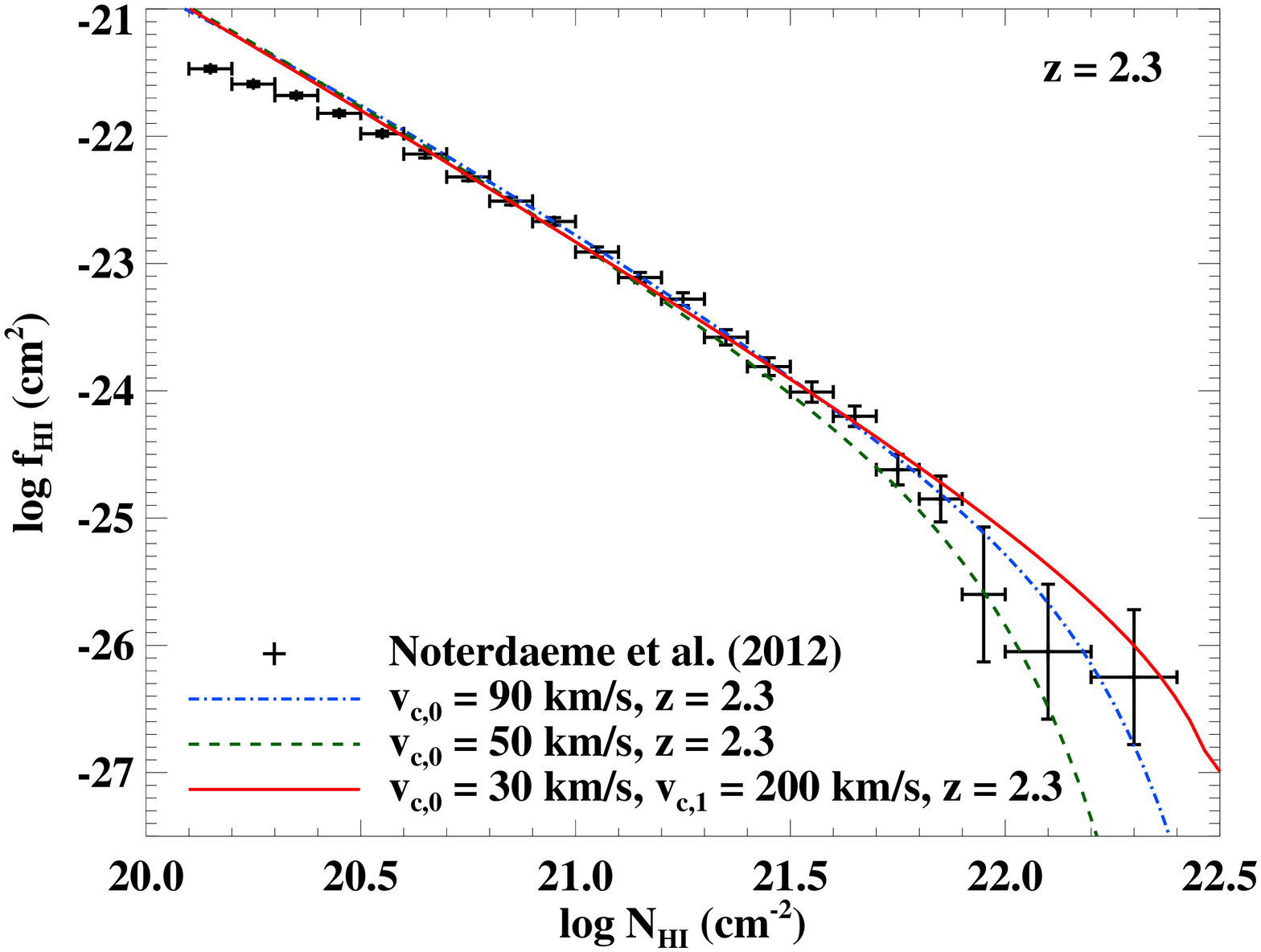}
   \end{center}
\caption{\textit{Top panel:} The column density distribution $f_{\rm HI}$ for the three models, at redshift $z \sim 0$, together with the fitting function of \citet{zwaan2005a}. \textit{Middle panel:} The $f_{\rm HI}$ at $z \sim 1$, with the data points from \citet{rao06}. \textit{Lower panel:} The $f_{\rm HI}$ at $z \sim 2.3$, with the data points from \citet{noterdaeme12}. The models are labelled by their cutoff velocities.}
   \label{fig:fHI}
  \end{figure}
  
  \begin{figure}
   \begin{center}
     \includegraphics[scale = 0.3, angle = 90]{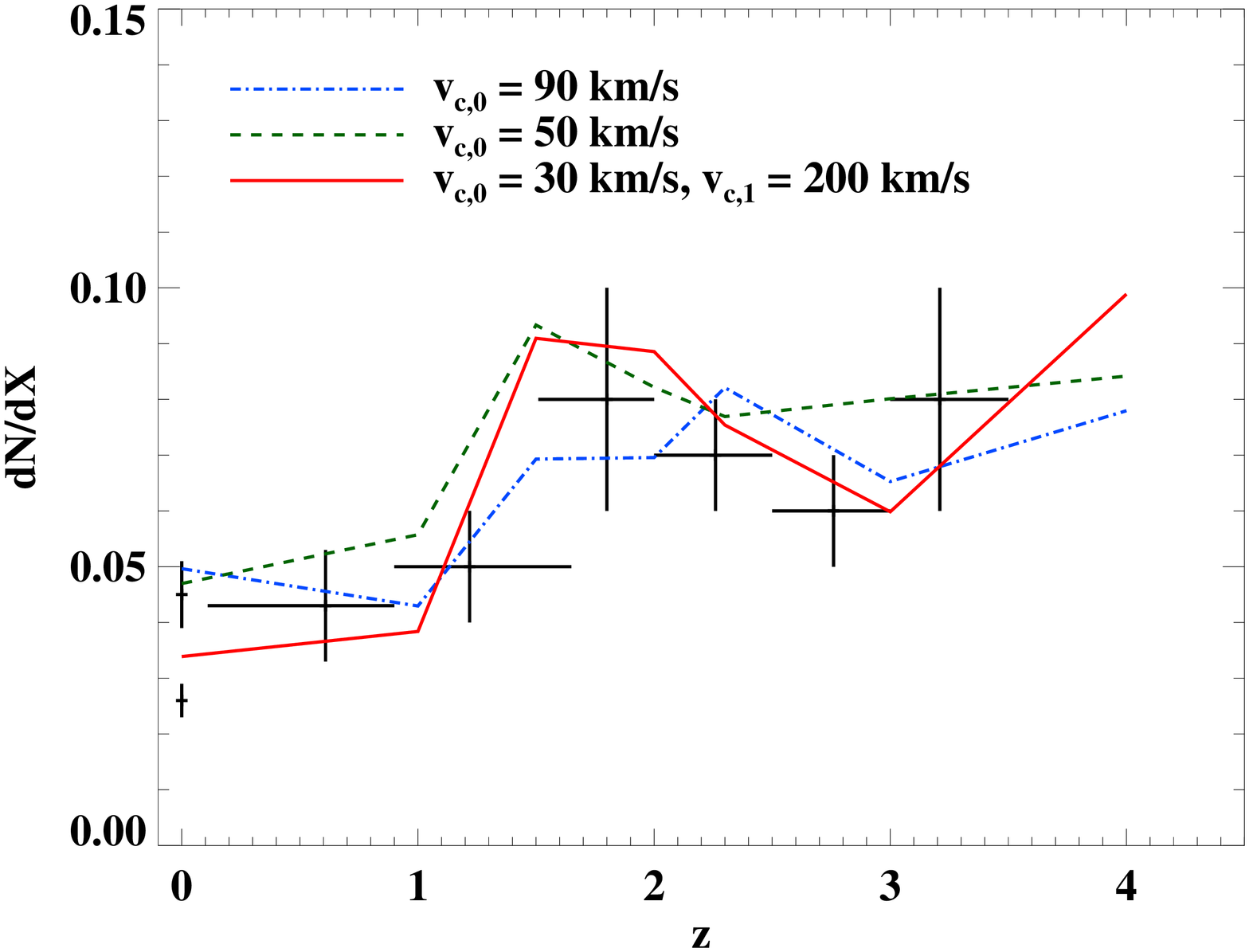} 
   \end{center}
\caption{The DLA incidence $dN/dX$ from  the observations of \citet{braun2012, zafar2013, zwaan2005a, rao06} and fitted by the three models considered. }
   \label{fig:dndx}
  \end{figure}

  \begin{figure}
   \begin{center}
   \includegraphics[scale = 0.3, angle = 90]{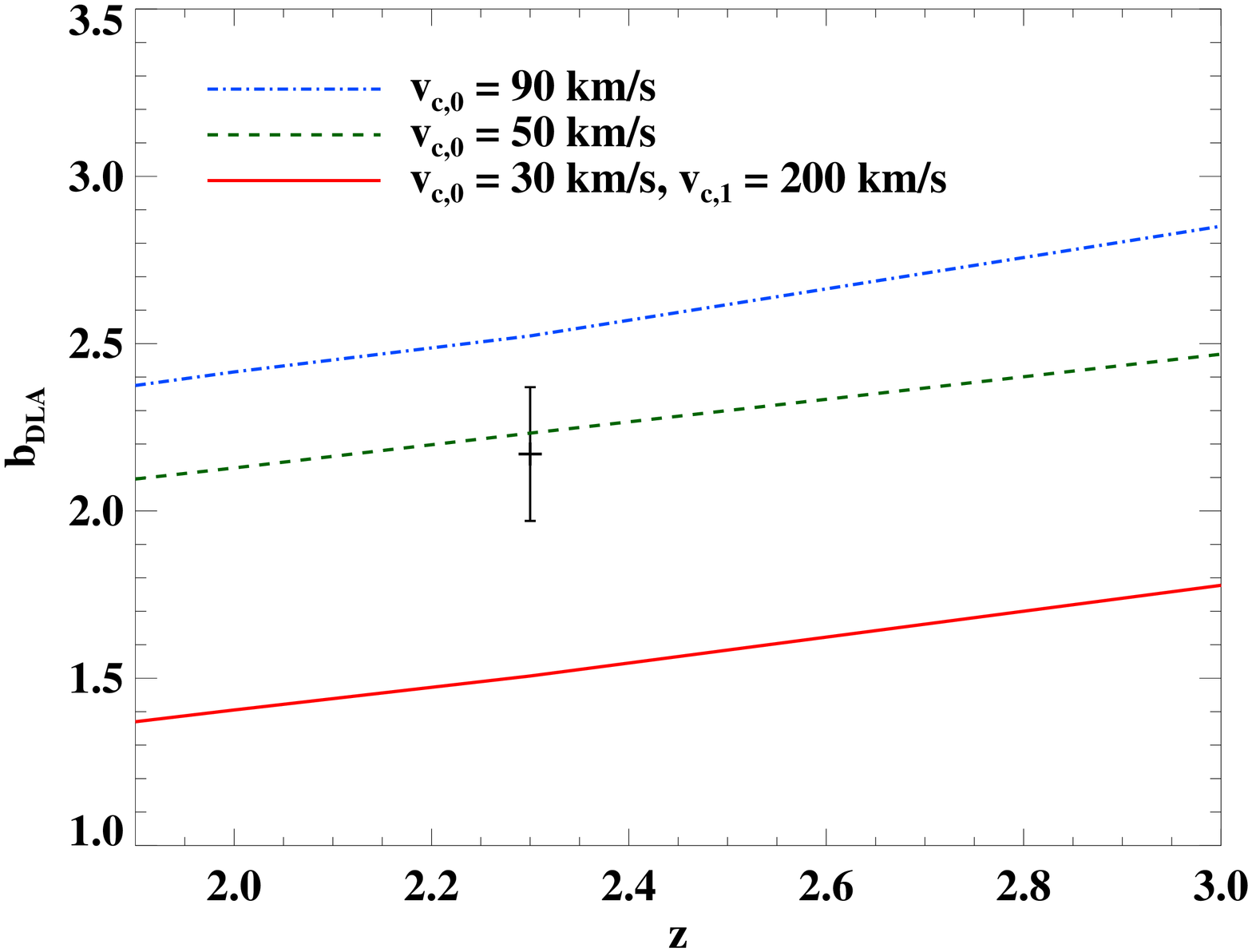} \includegraphics[scale = 0.3, angle = 90]{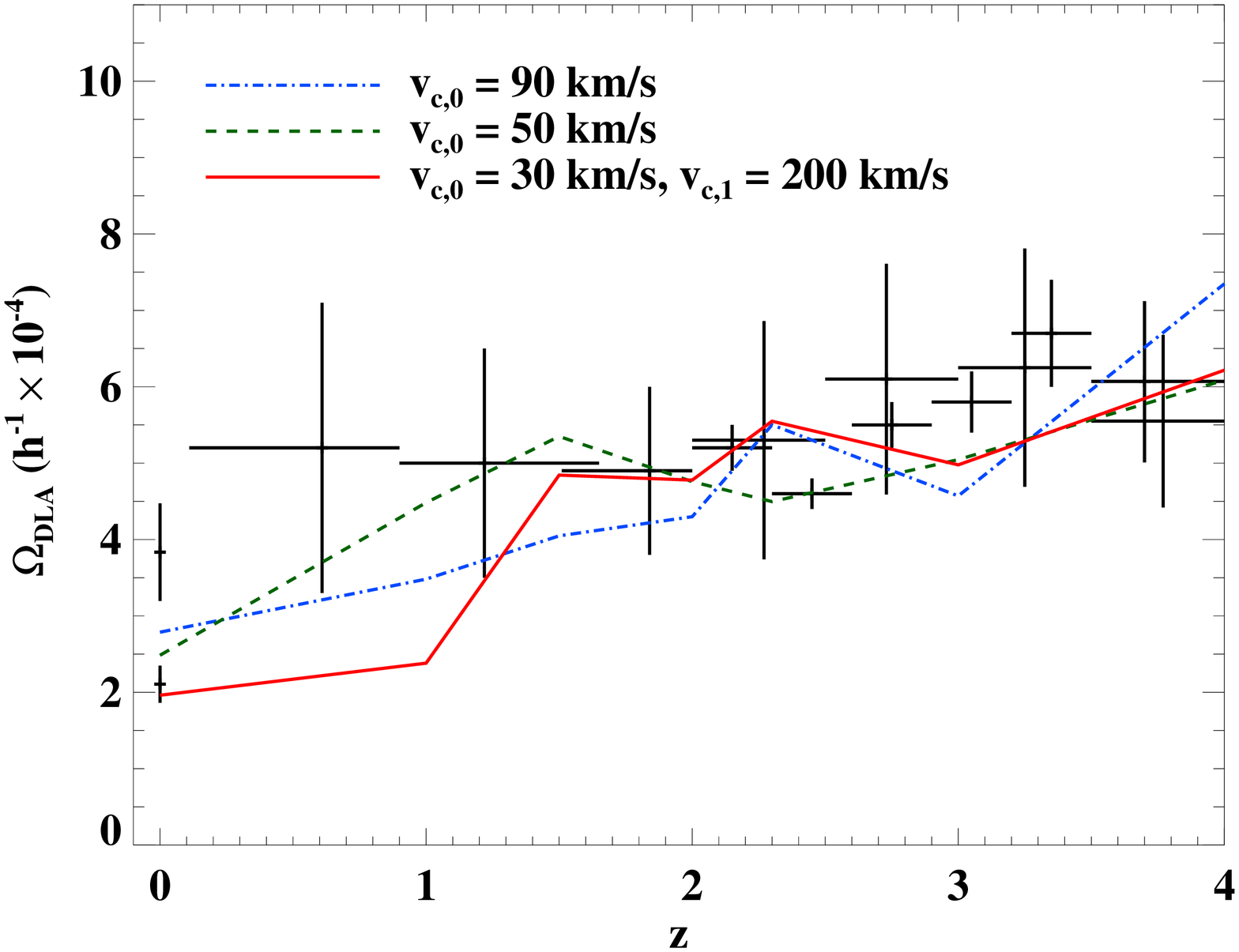}
   \end{center}
\caption{\textit{Top panel:} The predicted values of $b_{\rm DLA}$ from the three models, and the data point from \citet{fontribera2012}. \textit{Lower panel:} The values of $\Omega_{\rm DLA}$ from the three models, along with the data points \citep{zwaan05, braun2012, rao06, prochaska09, noterdaeme12, zafar2013}. }
   \label{fig:biasomDLA}
  \end{figure}

    \begin{figure}
   \begin{center}
    \hskip-0.25in \includegraphics[scale = 0.3, angle = 90]{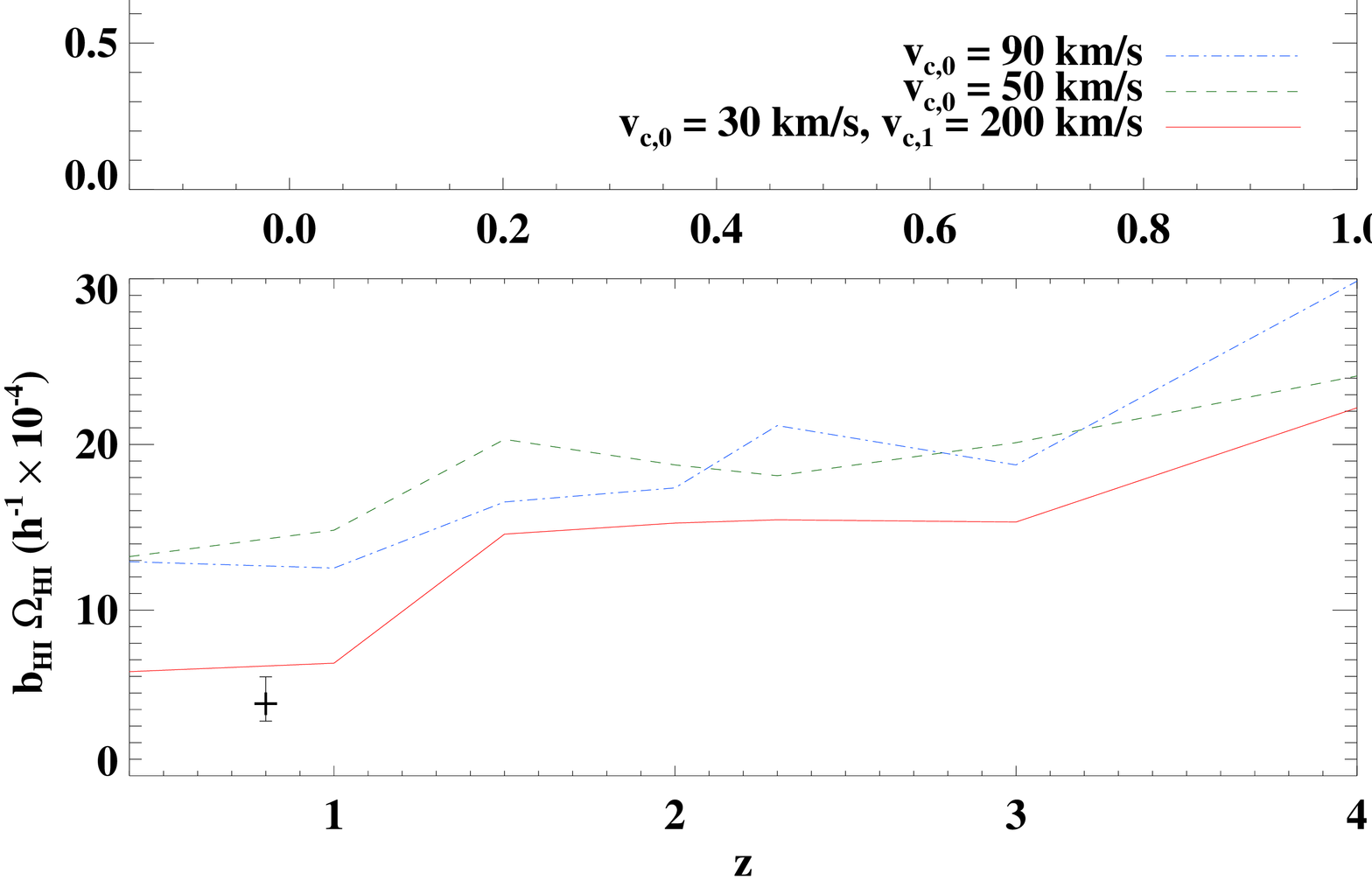}
   \end{center}
    \vskip-1.5in
\caption{{The low-redshift data, along with the model predictions. \textit{Top panel:} The bias $b_{\rm HI,gal}$ along with the $z \sim 0$ measurement \citep{martin12}.  \textit{Lower panel:} The product $\Omega_{\rm HI} b_{\rm HI}$ at all redshifts $z \sim 0-4$ as predicted by the models, along with the $z \sim 0.8$ measurement \citep{switzer13}.}}
   \label{fig:lowz}
  \end{figure}

 We use the mass prescription \eq{prescrip} with the HI profile given by \eq{rhodef} to model the observables, for all three choices of cutoffs: (a) $v_{c,0} = 90$ km/s, $v_{c,1} = \infty$ (b) $v_{c,0} = 50$ km/s, $v_{c,1} = \infty$ and (c) $v_{c,0} = 30$ km/s, $v_{c,1} = 200$ km/s. At each redshift, we vary the parameters $c_{\rm HI}$ and $\alpha$ to match the observed column density distribution $f_{\rm HI}$, where available, and the incidence rate of the DLAs, $dN/dX$. The quantities $\Omega_{\rm HI}$, $b_{\rm DLA}, b_{\rm HI, gal}$ and $\Omega_{\rm DLA}$ then arise as predictions of the model.   {Note that the calculated values of the bias $b_{\rm HI}$ are fixed by the cutoffs considered in the prescription, and do not depend on $c_{\rm HI}$ and $\alpha$. Hence, a different evolutionary scenario for $\alpha$ and $c_{\rm HI}$ would also lead to the same values of the calculated bias.}

 \begin{table}
\begin{center}
    \begin{tabular}{c  |c | c | c | c | c |c | c |}
    \hline
    \backslashbox{$v_{c,0}$}{$z$}
      & 0 & 1 & $1.5$ & $2$ & $2.3$ & $3$ & $4$   \\ \hline   
  90 km/s & 0.11 & 0.13  & 0.2 & 0.25 & 0.34 & 0.4 & 1.0 \\ 
  50 km/s & 0.09 & 0.13  & 0.2 & 0.21 & 0.22 & 0.3 & 0.5 \\ 
  30 km/s & 0.15 & 0.15  & 0.3 & 0.3  & 0.3  & 0.3 & 0.48 \\ \hline
    \end{tabular}
\end{center}
\caption{The values of $\alpha$, the overall normalization relative to cosmic of neutral hydrogen in the three models under consideration, chosen for consistency with the observed values of the incidence rate $dN/dX$ at redshifts 0 - 4, and the $f_{\rm HI}$ distribution, where available.}
 \label{table:alphas}
\end{table}

  \subsection{Fitting the models to the observations}
  
  We now use the models and the available data to constrain the free parameters. {In this work, we allow the parameters to vary independently at each redshift, thus making no assumptions \textit{a priori} about how they evolve with redshift. This ensures that we exploit the maximum freedom available within the models to draw conclusions which are robust.} Since the available observations are not all at the same set of redshift values, we choose a representative set of redshifts (0, 1, 1.5, 2, 2.3, 3, 4) at which the model outputs are generated. At $z = 0$,  $z = 1$, and 2.3, the parameters $c_{\rm HI}$ and $\alpha$ are fixed by the observed $f_{\rm HI}$. The $c_{\rm HI}$ determines the overall "slope" of the curve, and $\alpha$, the normalization, determines its height (this is discussed in detail in \citet{barnes2010}, see figure 7 of the paper) and hence both these parameters are constrained by the observed $f_{\rm HI}$. The value of $c_{\rm HI} = 25$ at $z \sim 2.3$ in both the DLA-based models \citep{barnes2010, barnes2014}, and is found to be higher ($c_{\rm HI} = 40$ and $c_{\rm HI} = 100$ respectively) at $z = 0,1$. For all the three redshifts,  $c_{\rm HI} = 60$ in the 21-cm based model. At $z = 1.5, 2, 3, 4$, where the observations of $f_{\rm HI}$ are not available, the parameters $c_{\rm HI}$ and $\alpha$ are chosen to match the observations of $dN/dX$ and $\Omega_{\rm DLA}$. The remaining observables then arise as the model predictions. Thus, the model free parameters are fixed to simultaneously match the observations of $f_{\rm HI}$ (where available), $\Omega_{\rm DLA}$ and  $dN/dX$ across redshifts 0-4. The values of $\alpha$ thus considered are indicated in Table \ref{table:alphas}, which indicate a trend of depleting HI gas towards lower redshifts.

   The $f_{\rm HI}$ distributions are plotted in Figs. \ref{fig:fHI} along with the available observational data from \citet[][a fitting function]{zwaan2005a}, \citet[][data points]{rao06} and \citet[][data points]{noterdaeme12}. The observed values of $dN/dX$ \citep{zwaan2005a, rao06, braun2012, zafar2013} are plotted in Fig. \ref{fig:dndx} and fitted by the three models.\footnote{The fitted value of $dN/dX$ at $z = 0$ is the average of the two measurements at this redshift, from \citet{zwaan2005a} and \citet{braun2012}.}
   
   The predicted bias $b_{\rm DLA}$ \citep{fontribera2012} and neutral hydrogen density parameter $\Omega_{\rm DLA}$ \citep{braun2012,  rao06, prochaska09, noterdaeme12, zafar2013} are plotted in Fig. \ref{fig:biasomDLA} along with the data points.\footnote{The \citet{zwaan05} measurement at $z = 0$ for atomic gas mass density has been corrected for by factor 0.81 for comparison to the measurements from DLAs, according to the results of \citet{zwaan2005a}.} It can be seen that the the 21-cm based prescription is a poor match to the observed value of DLA bias at $z \sim 2.3$, while the DLA-based predictions are closer to the observed value.
   
  The parameters $\Omega_{\rm HI}$ and  $b_{\rm HI}$ are then predicted by the models once the values of $c_{\rm HI}$ and $\alpha$ are fixed. \footnote{The predicted value of  $\Omega_{\rm HI}$ is always found to be greater than $\Omega_{\rm DLA}$. This is consistent with expectations because the $\Omega_{\rm DLA}$ contains contribution only from systems having $N_{\rm HI} > 10^{20.3}$ cm$^{-2}$, while the $\Omega_{\rm HI}$ measured from intensity mapping experiments contains contributions from the whole range of HI column densities.}
   The observation of  $b_{\rm HI}$ (at $z \sim 0$) from the ALFALFA survey\footnote{Above $10 h^{-1}$ Mpc, the bias parameter becomes independent of scale. The ALFALFA data provide the bias parameter as a function of scale; here, we plot the mean bias and its associated error at all scales $\gtrsim 10 h^{-1}$ Mpc as measured by this survey.}  \citep{martin12} is plotted in the top panel of Fig. \ref{fig:lowz} along with the predictions of the models.  At $z \sim 0$, the bias $b_{\rm HI,gal}$ measured by the ALFALFA survey is primarily sensitive to galaxies having HI masses above $10^{6} \ M_{\odot}$. Accordingly, at $z = 0$, we have neglected the contributions of $M_{\rm HI}$ masses below $10^6 \ M_{\odot}$ while calculating the values of $b_{\rm HI, gal}$ from the models. 
   
   {The model predictions for the product $\Omega_{\rm HI} b_{\rm HI}$ at all redshifts $z \sim 0-4$, along with  the intensity mapping measurement \citep{switzer13} at $z \sim 0.8$ are plotted in the lower panel of Fig. \ref{fig:lowz}. The model predictions at higher redshifts would be useful for comparing with observations in upcoming intensity mapping experiments.}

  The 21-cm based prescription ($v_{c,0} = 30$ km/s, $v_{c,1} = 200$ km/s) that we consider here, extended to account for the DLA observables,  is consistent with the majority of the low-redshift observations. It also matches  the DLA observables at higher redshifts except the high value of $b_{\rm DLA} \sim 2.17$ at redshift 2.3 (the model leads to $b_{\rm DLA} \sim 1.5$). The model also reasonably matches the constraint $\Omega_{\rm HI} b_{\rm HI}$ at $z \sim 0.8$ \citep{switzer13} and the bias $b_{\rm HI,gal}$ at low redshifts $z \sim 0$ \citep{martin12} as shown in Fig. \ref{fig:lowz}. 
  
  The DLA based prescriptions, on the other hand, lead to values of $b_{\rm HI}$ and $\Omega_{\rm HI} b_{\rm HI}$  which are higher than observed at $z \sim 0-1$. However, these prescriptions are consistent with the high value of $b_{\rm DLA}$ at $z \sim 2.3$.

\section{Discussion}

As we have seen, two independent observational techniques are used to constrain the distribution and  evolution of HI in the post-reionization universe --- the DLA observations and the 21-cm based measurements. These two observational techniques have, in turn, been typically associated with their own theoretical and analytical prescriptions. {{While the two sets of analytical techniques have been applied separately to the corresponding observations in the existing literature, here we attempt to connect these approaches to take into account all the available data, in a common framework from both these sets of measurements simultaneously.}} This is particularly relevant because both these observational probes --- DLAs as well as 21-cm based measurements --- are sensitive to the high-column density neutral gas, and hence also important probes of large-scale structure formation.

In this paper, we have attempted to reconcile the 21-cm and the DLA based prescriptions to describe the characteristics of HI host haloes at various redshifts in the post-reionization universe. We have summarized the existing models for the 21-cm and the DLA based prescriptions for assigning HI to dark matter halos. We have modified and extended the existing 21-cm based model to account for the DLA observations, and also extended the DLA-based models to all redshifts 0-4. In each case, and at every redshift, we have calibrated the model free parameters by matching the available observations. Our main findings may be summarized as follows:

 \begin{enumerate}
  \item A physically motivated, 21-cm based prescription ($v_{c,0} = 30$ km/s, $v_{c,1} = 200$ km/s), in combination with a halo profile for the distribution of HI, provides a good fit to the majority of the available data. It matches the  observed low value of the HI bias at $z \sim 0$,  and of the product $\Omega_{\rm HI} b_{\rm HI}$ at $z \sim 1$. The model is also consistent with with the DLA observations of $f_{\rm HI}$, $\Omega_{\rm DLA}$ and $dN/dX$ at $z \sim 0-4$. However, at $z \sim 2.3$, the model leads to a lower value of DLA bias $b_{\rm DLA}$ than is observed. 
  
  \item The two DLA-based models, having cutoffs of $v_{c,0} = 50$ and 90 km/s respectively, are consistent with the measurement of $b_{\rm DLA}$ at $z \sim 2.3$. The clustering measurement, therefore, requires the lower cutoff in the sampling of the halo mass function to be close to 50-90 km/s. It also favours the presence of a very high (or the absence of a)  high mass cutoff. Both these factors suggest that the observed DLA bias at $z \sim 2.3$ is well reproduced if the neutral hydrogen in shallow potential wells is depleted, thus suggesting the possibility of very efficient stellar feedback \citep{barnes2014}. However, it is difficult to reconcile models having such strong feedback with the low-redshift observations of HI bias and $\Omega_{\rm HI} b_{\rm HI}$.
  {{
  \item   This highlights a tension between the DLA bias and the 21-cm measurements, unless there is a significant change in the nature of HI-bearing systems across redshifts 0-3, i.e. from lower-mass haloes at low redshifts to more massive haloes at higher redshifts. It is important to note that this result holds irrespective of how the free parameters $\alpha$ and $c_{\rm HI}$ evolve with time, and thus directly constrains the maximum and minimum masses of the haloes under consideration.
   }}
  \item This also has implications for the measured 21-cm intensity fluctuation power spectrum. Taken together, the observed bias measurements at $z \sim 0$ and $z \sim 2.3$ suggest an almost factor of two evolution in the bias between redshifts 0-3, which alone implies a factor 4 evolution in the measured power spectrum of intensity fluctuations.
  \end{enumerate}
  
  These findings can be related to other lines of investigation into the nature of high-redshift DLA hosts.
  The results of the 21-cm based model lead to a DLA bias ($b_{\rm DLA} \sim 1.5$ at redshift $z \sim 2.3$) which is lower than the observed value. The model suggests that DLAs are hosted by faint dwarf galaxies at high redshift ($z \sim 2.3$). This finding is consistent with the observed lack of nearby high-luminosity galaxies at high redshifts \citep{cooke2015}. It has been suggested \citep{fumagalli2015} that the DLAs are hosted by faint dwarf galaxies which are either isolated or clustered with more massive galaxies. The observed high value of bias $b_{\rm DLA}$ may be explained if the DLAs arise from dwarf galaxies which are satellites of massive Lyman-break galaxies \citep{fontribera2012}. However, the lack of bright Lyman-break galaxies in the vicinity of DLAs \citep{fumagalli2015} suggests that DLAs may be isolated dwarf Lyman-break galaxies, which would lead to a much lower bias than observed at $z \sim 2.3$. The low star formation rates \citep[$0.09 - 0.27 M_{\odot}$ yr$^{-1}$;][]{fumagalli2015} associated with these systems typically correspond to host haloes of masses $\lesssim 10^{11} M_{\odot}$ at $z \sim 2$ \citep[e.g.,][]{behroozi2013}.  The results of cosmological simulations also support this view since DLAs arise in host haloes of masses $10^9 - 10^{11} M_{\odot}$ at redshift $z \sim 3$ \citep{pontzen2008, tescari2009, fumagalli2011, cen2012, voort2012, bird2013, rahmati2014}.\footnote{The incidence rate for the DLAs as a function of host halo mass, also shows a peak around host halo masses of $M \sim 10^{9.5} - 10^{12} M_{\odot}$ at $z \sim 2.3$ in the 21-cm based prescription, consistent with previous findings of \citet{barnes2010} which point to DLAs being hosted by dark matter haloes with virial velocities in the range 35-230 km/s.} It is also found in simulations \citep{dave2013} that at these redshifts, only half the cosmic HI resides in $M_{\rm HI} > 10^9 M_{\odot}$ systems, with a very small fraction in $M_{\rm HI} > 10^{10}  M_{\odot}$ systems.  
  Hence, the 21-cm based model, while inconsistent with the high-redshift bias measurement, is indeed consistent with the recent results of imaging and other surveys of high-redshift DLAs. 

In future work, it would be useful to compare our analytical results with the semianalytical and simulation studies \citep[e.g.,][]{bird2013, rahmati2014, cen2012, tescari2009}. This would, in turn, also have important consequences for determining the 21-cm intensity fluctuation power spectrum, and hence better understanding the distribution and evolution of neutral hydrogen in the post-reionization universe.

 \section*{Acknowledgements}
 
 HP acknowledges support from the SPM research grant of the Council of Scientific and Industrial Research (CSIR), India. We thank Varun Bhalerao, Jayaram Chengalur, Romeel Dav\'{e}, Nissim Kanekar, Roy Maartens, Sebastian Seehars, R. Srianand, Kaustubh Vaghmare and Francisco Villaescusa-Navarro for useful discussions and comments on the manuscript. We thank the anonymous referee for useful comments that improved the content and presentation of the paper.

\bibliographystyle{mn2e} 
\def\aj{AJ}                   
\def\araa{ARA\&A}             
\def\apj{ApJ}                 
\def\apjl{ApJ}                
\def\apjs{ApJS}               
\def\ao{Appl.Optics}          
\def\apss{Ap\&SS}             
\def\aap{A\&A}                
\def\aapr{A\&A~Rev.}          
\def\aaps{A\&AS}              
\def\azh{AZh}                 
\def\baas{BAAS}
\def\jcap{JCAP}
\def\jrasc{JRASC}             
\def\memras{MmRAS}
\def\na{New Astronomy}
\def\nat{Nature}
\def\mnras{MNRAS}             
\def\pra{Phys.Rev.A}          
\def\prb{Phys.Rev.B}          
\def\prc{Phys.Rev.C}          
\def\prd{Phys.Rev.D}          
\def\prl{Phys.Rev.Lett}       
\def\pasp{PASP}               
\def\pasj{PASJ}
\def\physrep{Phys. Repts.}
\def\qjras{QJRAS}             
\def\skytel{S\&T}             
\def\solphys{Solar~Phys.}     
\def\sovast{Soviet~Ast.}      
\def\ssr{Space~Sci.Rev.}      
\def\zap{ZAp}                 
\let\astap=\aap
\let\apjlett=\apjl
\let\apjsupp=\apjs

\bibliography{mybib}

\end{document}